\documentclass[preprint]{elsarticle}
\usepackage{lineno,hyperref}

\usepackage[version=3]{mhchem} 
\usepackage{upgreek} 

\usepackage{xcolor}
\newcommand{\hl}[1]{\textcolor{black}{#1}}
\modulolinenumbers[5]

\journal{Journal of Colloids and Interface Science}

\begin{document}

\begin{frontmatter}

\title{How water wets and self-hydrophilizes\\ nanopatterns of physisorbed hydrocarbons}


\author[uc,max]{Diego Díaz}
\author[hamburgopol]{Ole Nickel}
\author[uc]{Nicolás Moraga}
\author[uc]{Rodrigo E. Catalán}
\author[uc]{María José Retamal}
\author[uc]{Hugo Zelada}
\author[uc]{Marcelo Cisternas}
\author[hamburgopol]{Robert Mei{\ss}ner}
\author[hamburgofun,desy,uhamburgo]{Patrick Huber\corref{mycorrespondingauthor}}
\ead{patrick.huber@tuhh.de}
\author[utfsm]{Tomás P. Corrales\corref{mycorrespondingauthor}}
\ead{tomas.corrales@usm.cl}
\author[uc,cien]{Ulrich G. Volkmann\corref{mycorrespondingauthor}}
\ead{volkmann@fis.puc.cl}

\cortext[mycorrespondingauthor]{Corresponding authors}

\address[uc]{Instituto de Física, Pontificia Universidad Católica de Chile, Santiago 7820436, Chile}
\address[hamburgopol]{Hamburg University of Technology, Institute of Polymers and Composites, 21073 Hamburg, Germany}
\address[hamburgofun]{Hamburg University of Technology, Institute for Materials and X-Ray Physics, 21073 Hamburg, Germany}
\address[desy]{Deutsches Elektronen-Synchrotron DESY, Centre for X-Ray and Nano Science CXNS, 22603 Hamburg, Germany}
\address[uhamburgo]{University of Hamburg, Centre for Hybrid Nanostructures CHyN, 22607 Hamburg, Germany}
\address[utfsm]{Departamento de Física, Universidad Técnica Federico Santa María, Valparaiso 2390123, Chile}
\address[cien]{Centro de Investigación en Nanotecnología y Materiales Avanzados (CIEN-UC), Pontificia Universidad Católica de Chile, Santiago 7820436, Chile}
\address[max]{Max Planck Institute for Polymer Research, Ackermannweg 10, D-55128, Mainz, Germany}

\begin{abstract}
{
\textit{Hypothesis}\\
Weakly bound, physisorbed hydrocarbons could in principle provide a similar water-repellency as obtained by chemisorption of strongly bound hydrophobic molecules at surfaces.\\\\
\textit{Experiments}
\\
Here we present experiments and computer simulations on the wetting behavior of water on molecularly thin, self-assembled alkane carpets of dotriacontane (\ce{$n$-C32H66} or C32) physisorbed on the hydrophilic native oxide layer of silicon surfaces during dip-coating from a binary alkane solution. By changing the dip-coating velocity we control the initial C32 surface coverage and achieve distinct film morphologies, encompassing homogeneous coatings with self-organized nanopatterns that range from dendritic nano-islands to stripes.
\\\\
\textit{Findings}
\\
These patterns exhibit a good water wettability even though the carpets are initially prepared with a high coverage of hydrophobic alkane molecules. Using in-liquid atomic force microscopy, along with molecular dynamics simulations, we trace this to a rearrangement of the alkane layers upon contact with water. This restructuring is correlated to the morphology of the C32 coatings, i.e. their fractal dimension. Water molecules displace to a large extent the first adsorbed alkane monolayer and thereby reduce the hydrophobic C32 surface coverage. Thus, our experiments evidence that water molecules can very effectively hydrophilize initially hydrophobic surfaces that consist of weakly bound hydrocarbon carpets.}
\end{abstract}

\end{frontmatter}


\section{Introduction}
The wetting and spreading of liquids on planar silicon surfaces and in porous silicon, with and without native silicon oxide layers, plays a crucial role in a large variety of technologies ranging from nanofluidics via energy storage to energy harvesting and the design of new materials via 3-D printing \cite{deGennes2004CapillarityPhenomena, Bonn2009, Ueda2013, Huber2015, Ocier2020}. Also key fundamental insights on wetting and dewetting could be achieved by the study of silicon-silica surfaces by systematically tuning the liquid-solid interaction by chemical surface grafting \cite{deGennes2004CapillarityPhenomena, Checco2010, Checco2014,Lessel2015,Schellenberger2016HowSurfaces, Shou2019}. 

Whereas hydrogen-terminated silicon is hydrophobic, the formation of native oxide layers and subsequent hydroxylation leads to polar, hydrophilic surfaces \cite{deGennes2004CapillarityPhenomena}. Since oxidation of silicon is unavoidable under standard atmospheric conditions, chemical surface grafting, e.g. grafting of strongly bound linear hydrocarbon chains by silanization, is a frequently used technique to make silicon with native silica surfaces hydrophobic \cite{Wasserman1989, deGennes2004CapillarityPhenomena,Steinrueck2014, Lessel2015,Gruener2016, Gruener2019}. An example is the silanization of soot-templates surfaces, which have been shown to produce transparent and superhydrophobic surfaces \cite{Deng2012}. Although these surfaces are self-cleaning, they dynamically change their wetting properties when interacting with ethylene glycol and ionic liquids \cite{Wong2020MicrodropletSelf-Cleaning}. 

There are several chemisorbed hydrophobic surfaces that adapt in response to liquid interaction, i.e., they reversibly change their conformation \cite{Butt2018AdaptiveWetting}. However, irreversible rearrangements of physisorbed hydrophobic molecules that adapt to water have not been explored so far, even though they are of potential relevance for liquid-functionalized porous solids and surfaces \cite{Wong2011, Huber2015,Gang2020,Peppou-Chapman2020, Peppou-Chapman2021} as well as the broad field of physico-chemical processes and fluid transport in porous media, e.g. in nanofluidics, enhanced oil recovery in unconventionally tight oil reservoirs and membrane separation processes. {Weakly bound physisorbed hydrocarbon layers can potentially alter wettability and thus hydraulic permeabilities and spontaneous imbibition processes \cite{Gruener2009, Huber2015, Gruener2016, Gruener2016a, Vincent2016,Gruener2019, Monroe2021}.} {Also the changes of the wettability of graphitic interfaces by airborne physisorbed hydrocarbon carpets has recently been in the focus of active research with regard to an apparent increase in the hydrophobicity of graphene/water interfaces \cite{Li2013, Kozbial2014, Terzyk2019}.}

Motivated by our own findings on the spontaneous formation of ultra-thin physisorbed nanopatterns of n-alkane films with distinct, preparation-dependent morphology on silicon-silica surfaces, we present here a study on the wetting of weakly bound, i.e. physisorbed, molecular films by water. In the past, the self-assembly of n-alkanes has been studied on dry surfaces, such as graphite \cite{Hansen2004IntramolecularSimulations}, gold \cite{Soza2006EllipsometricSurfaces} and silicon\cite{Volkmann2002High-resolutionSurface}. For silicon (111), with a native oxide layer of roughly 1 nm thickness, \ce{$n$-C32H66} (C32 in the following) forms a parallel bilayer of molecules followed by several monolayers of standing molecules oriented perpendicular to the surface. This growth model has been reported for vapor deposited and dip-coated films by employing surface-sensitive x-ray scattering methods\cite{DelCampo2009}, atomic force microscopy (AFM) \cite{Trogisch2005AtomicSurface} and ellipsometry \cite{Volkmann2002High-resolutionSurface}. \hl{Molecular Dynamics simulations corroborate this peculiar multilayer structure \cite{Mo2004StructureSimulations,Gutierrez-Maldonado2017AccessingSurface} and trace it to a subtle interplay of intermolecular interactions of the rod-like molecules and their interaction with the oxidic substrate.} 

In a recent study, we have shown that C32 can form controllable nanopatterns on silicon substrates by dip-coating the sample at different velocities in binary alkane solutions \cite{Corrales2014SpontaneousStripesb}. By changing the pulling velocity $v$, we have explored two distinct coating regimes: i) an evaporation regime and ii) a film entrainment regime, see Fig. \ref{fig:dipcoating}. These two regimes are separated by a critical velocity $v_c$, that depends on the solvent. In Ref.~\citenum{Corrales2014SpontaneousStripesb} we used n-heptane (\ce{$n$-C7H16}) to dissolve C32 molecules and have found that at a fixed solution concentration (0.75 {g/L}), a single perpendicular layer of molecules is formed over a homogeneous coverage of 2 flat lying molecular layers covering the entire substrate for all pulling velocities. Moreover, by changing the pulling velocity, the surface coverage and pattern morphology of the perpendicular layer can be controlled, see Fig.~\ref{fig:dipcoating}. For small $v$, seaweed-like and dragonfly-shaped molecular islands are observed. For a large $v$, stripes parallel to the withdrawal direction emerge. These have lengths of a few hundred micrometers and a few micrometer lateral separation.

Since native silicon oxide is hydrophilic and the C32 molecule hydrophobic, we can fabricate a well-defined, self-organised mesoscale wetting heterogeneity. In the following we employ this controllable nano-pattering process to study the wetting of water droplets on such chemically heterogeneous surfaces. Hydrophobic molecules covering a hydrophilic background should affect the macroscopic contact angle. To that end we present contact angle measurements of water droplets on such n-alkane nanopatterns and relate them with Molecular Dynamics computer simulations. The wetting changes are rationalised by considering the heterogeneous wetting geometry (Cassie model) as well as the fractal dimension of the formed C32 patterns and their dynamic restructuring when interacting with water.

\section{Experimental Section}

\subsection{Sample Preparation}

\subsection{Contact Angle Measurements}
We perform contact angle measurements using an in-house built sessile drop setup which is composed of a 5.32x 1.89\,cm$^2$ steel-plate, on which the Si substrates are placed. Beneath the plate, a goniometer is installed to rotate the sample. The goniometer plus steel-plate are coupled to a system of platforms moved by micrometer screws. This allows us to move the sample in three dimensions plus rotation. Ultra-pure water droplets (Merck, 18\,M$\Omega$cm) of 2\,$\upmu$L were placed on the substrate centers with different coverages using a micropipette (Optipette). The micropipette was coupled using laboratory clamps to a labjack which allowed us to gently place the drop on the surface. The camera utilised for contact angle measurements was a digital microscope (Digi-Microscope 0.3 MP, 50-500X) and the calibration established for droplet images was 14\,$\upmu$m/px. All analysed images had sizes of 640x480\,px. Contact angles were calculated by fitting a circle to the droplet contour and measuring its curvature radius and height. From the measured height and radius of the drop, we calculate the contact angle using {the equation taken from \cite{schmieschek2011,Huang2007}}: 
\begin{equation}
\theta^{*}=\arctan \sqrt{\frac{h}{2r-h}} 
\end{equation}
Where $\theta^*$ is the measured apparent contact angle, $h$ is the height of the drop and $r$ is the fitted circle radius.
\subsection{SEM and AFM measurements of nanopatterns}
Coverages and morphology were measured by scanning electron microscopy (SEM). {All SEM images are taken at 3000x magnification and 25\,kV beam energy  (LEO VP1400 operated in high vacuum).} Atomic Force Microscopy (AFM)  is operated in tapping mode, both in air and liquid conditions using a NanoWizard 3 BioScience AFM (JPK Instruments). The resonance frequency of the AFM cantilevers (Nano and More, USA) was 300\,kHz with a constant force of 50\,N/m. 

\subsection{Image processing}
Scanning electron microscopy (SEM) images of all samples are binarised using imageJ. After {image binarization}, the number of white pixels that represent the C32 perpendicular layer, are counted and divided by the total pixels to obtain the surface coverage percent. Meanwhile, the fractal dimension of our samples was measured using a box counting method implemented in ImageJ. The fractal dimension is defined here as:
\begin{equation}
D_{\rm F}=\frac{\log N}{\log \epsilon }
\end{equation}
Where N is the number of boxes used to cover our nanopatterns and $\epsilon$ is the size of the boxes.
\subsection{Molecular dynamics simulations}

All MD simulations were performed with the Large-scale Atomic/Molecular Massively Parallel Simulator (LAMMPS) \cite{Plimpton1995} in the version dated 3 March 2020 utilizing a recently published extension of the OPLS force field to describe accurately long hydrocarbon chains \cite{Siu2012} in combination with the TIP3P water model \cite{Jorgensen1983}.

Interactions between the silica surface, water and hydrocarbon chains are described using parameters as given in Butenuth \textit{et al.} \cite{Butenuth2012}.
Bulk silica interactions are described by a modified force field based on a Morse stretch potential \cite{Demiralp1999} as described in detail in Mei{\ss}ner \textit{et al.} \cite{Meissner2014}.

Silicon spontaneously forms an amorphous silicon dioxide layer under ambient conditions \cite{Cole2007}.
Thus, an amorphous deprotonated silica slab with surface areas ranging between 120\,-\,150\,nm\textsuperscript{2} (depending on the water droplet size and geometry) and a silanol group density of 3\,OH/nm\textsuperscript{2} was employed as a model system for the oxidised and 1.5~nm thick amorphous SiO$_2$ layer on the surface of the silicon wafer substrate. 
The surface charge of silica depends strongly on pH, ionic strength and in case of colloidal particles on the diameter \cite{Behrens2001,Barisik2014}.
The surface charge density at pH 7.0 amounts to about 0.55\,$e\cdot$nm\textsuperscript{-2} (corresponding to 0.5 Si-O$^-$ groups per nm\textsuperscript{-2}) \cite{Behrens2001}.
This surface charge density is set by deprotonation of randomly chosen silanol terminal groups (cf. Mei{\ss}ner \textit{et al.} \cite{Meissner2014}).
To ensure the charge neutrality of the whole system, sodium counter-ions not interacting with water and hydrocarbon chains were inserted on the opposite side of the surface where no wetting occurs, at random positions 3.5\,nm under the slab, and limited to their initial positions.

Only $x$ and $y$ dimensions are periodic for which a k-space solver with an accuracy of 10\textsuperscript{-6} and corrections for slab geometries is used \cite{Yeh1999}.
In all simulations the timestep was 1\,fs and the temperature was set to 298\,K by coupling the system to a heat bath with a Nos\'e-Hoover thermostat \cite{Shinoda2004}.
Bonds and angles of water molecules and bonds in hydroxyl groups on the surface are held rigid with SHAKE {algorithm} \cite{Ryckaert1977}.
The cutoffs for Morse, van der Waals and the Coulombic part of the real space are 0.9, 1.0 and 1.2\,{nm} respectively.
Initially hydrocarbon chains (n-\ce{C32H66}) with their approximate assumed adsorption configuration (perpendicular and/or two to three planar layers) are placed on the surface and the resulting structure was minimised energetically.

The system was then brought into thermal equilibrium at 300\,K within a few nanoseconds and either a spherical or a cylindrical water droplet, the latter flat on the surface, was added after reaching a steady state.
The amount of water in the droplet and the column ranged between 4000 and 12000 H\textsubscript{2}O molecules.
To measure the contact angle, a circle is fitted to the isovalue of a two-dimensional density profile, averaged over several nanoseconds, at which the density is approximately a quarter of the bulk density of water {\cite{Woch2016,Jiang2019}}.
From the angle between the surface and the tangent at which the circle crosses the surface plane the contact angle was estimated.

\section{Results and discussion}
\begin{figure}[h]
\centering
\includegraphics[width=\columnwidth]{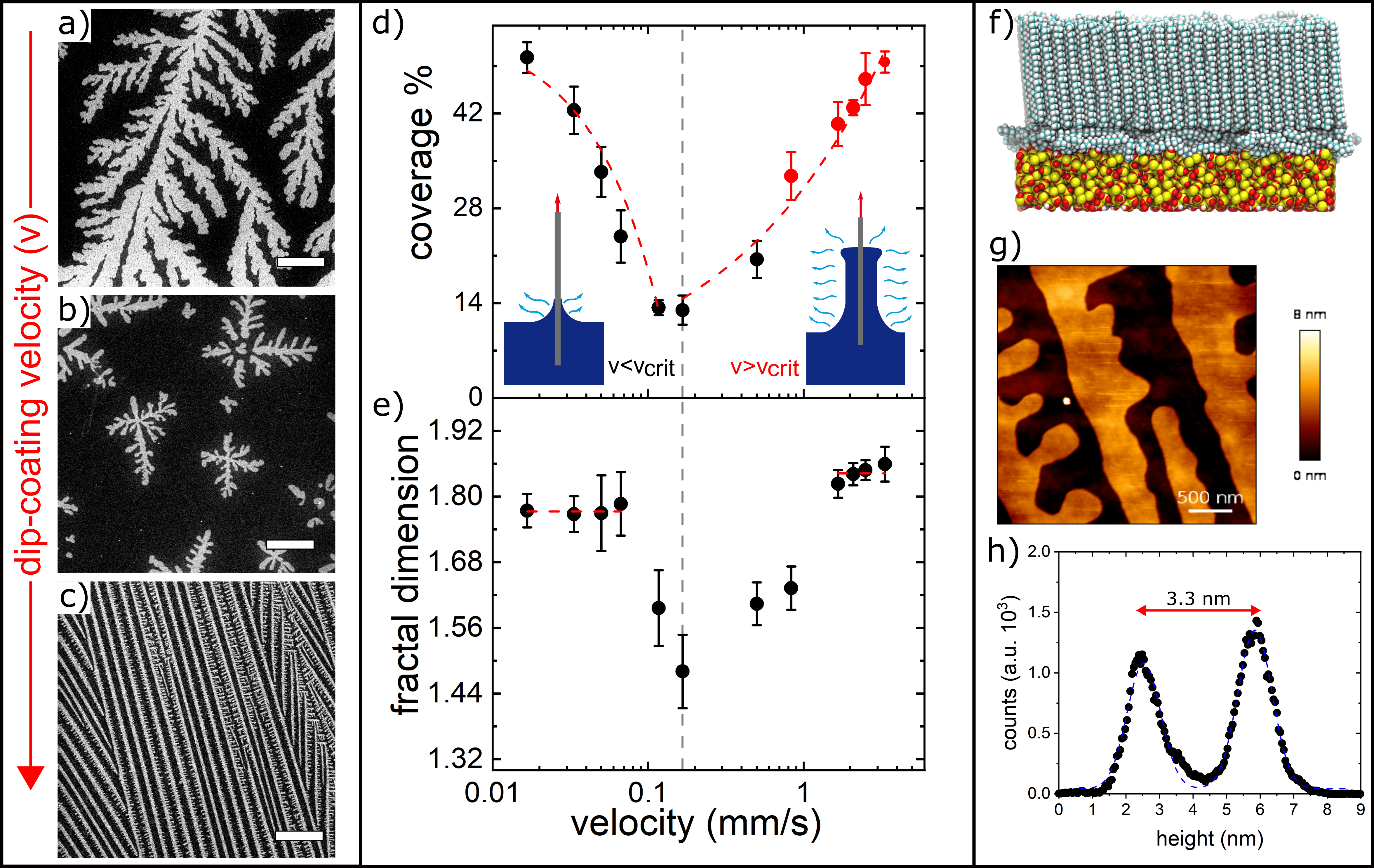}
  \caption{\textbf{Velocity-dependent dip-coating of silicon substrates in binary hydrocarbon solutions results in ultra-thin C32 hydrocarbon films with distinct morphologies.} {Top-view SEM micrographs of C32 samples dip-coated at 3 velocities: a) 0.017, b) 0.17 and c) 3.3 mm/s. The white scale bar represents 10 $\upmu$m. d) Surface coverage and e) fractal dimension of nanopatterns as determined from SEM micrographs. The insets depict the evaporation ($v<v_{\rm crit}$) and entrainment regime ($v>v_{\rm crit}$).  f) Dip-coating at all velocities result in distinct nanopatterns consisting of a parallel molecular bilayer followed by a single standing layer perpendicular to the surface, which is supported on a native silica layer of 1.5\,nm. g) AFM image of a sample prepared at 3.3 mm/s and h) its corresponding height histograms. The difference between the height histogram maxima, i.e., the difference between the dark valleys and the bright monolayer surfaces, correspond to a height typical of a perpendicular C32 monolayer.}}
  \label{fig:dipcoating}
\end{figure}

{\subsection{Alkane film morphology and fractal dimension analysis}}
After preparing samples in duplicate by dip-coating at a fixed velocity, one of the samples is taken to the scanning electron microscope (SEM) and the other twin-sample to the contact angle experiment. SEM images reveal the influence of the pulling velocity on the surface coverage and the morphology of the single perpendicular layer of molecules. In Figure\ref{fig:dipcoating}a-c we show representative electron micrographs {for samples prepared at 3  different pulling velocities}. All images are 60x60\,$\upmu$m$^2$ in size and show the differences in mesoscopic assembly of the C32 film, in particular the distinct patterns of the perpendicular layer. 
 To quantify the coverage,  all prepared samples were measured with SEM. Three samples per velocity were taken to SEM. For each sample, 5 corner regions and the center are imaged, resulting in 15 SEM images for each preparation velocity. All images are binarised and white pixels are counted over these regions to obtain an average surface coverage for each preparation velocity. SEM images, for all samples prepared between 0.017 to 3.3 mm/s, are shown in supplementary information (Fig. S1). To characterise the morphology of the nanopatterns, we determined their fractal dimension $D_F$. In Fig. \ref{fig:dipcoating}d-e we show the surface coverage and fractal dimension versus dip-coating velocity $v$.

The surface coverage and fractal dimension reach a minimum at 0.17~mm/s (Fig. \ref{fig:dipcoating}d-e). {The surface coverage decreases linearly between 0 and 0.12 mm/s. At 0.17 mm/s, the surface coverage reaches a minimum and starts growing following a power law.} {The common minimum of both coverage and fractal dimension is marked as a grey dashed line in figure \ref{fig:dipcoating}.}

These two regimes are related to distinct C32 deposition modes. Below {a dip-coating velocity of} 0.17~mm/s, no liquid film is pulled from the solution and the material is deposited by molecular diffusion and evaporation at the triple phase contact line of the meniscus (substrate/bulk solution/vapor).{ We refer to this deposition mode as the evaporation regime} (see Ref. \citenum{Corrales2014SpontaneousStripesb}) . {Above a dip-coating velocity of 0.17 mm/s, a liquid film of solution is pulled from the bulk solution, covering the Si-substrate. This deposition regime is referred to as the entrainment regime. The thickness of the liquid film in the entrainment regime follows a Landau-Levich power law,  scaling like $\sim v^{0.35}$ \cite{Landau1942,Tewes2019}. Both coating regimes are depicted in the inset of Fig. \ref{fig:dipcoating}d. The final film coverage is directly influenced by these coating regimes, as described in  \cite{Corrales2014SpontaneousStripesb}. }

{Visual inspection of} the nanopatterns exhibit a {morphological} change from seaweed-like structures via compact dendritic island structures to stripes. This structure formation results from a complex interplay of {evaporation}, crystallization, diffusion and hydrodynamic flow, similarly as has been observed for other dip-coated organic films \cite{Rogowski2010}. We analyze this {behaviour} by considering the fractal dimensions $D_F$ of the patterns. The fractal dimension for velocities up to 0.06~mm/s, where seaweed-like structures are found, is $D_F \sim 1.7$, as seen in Figure \ref{fig:dipcoating}e. Thus, the C32 dendrites have a $D_{\rm F}$ in between the one typical of a Koch snowflake ($D_{\rm F}=$1.26 \cite{Backes2012}) and an Euclidean square ($D_{\rm F}=$2.0). In fact, a value of $D_F$=1.7 is characteristic of growth governed by diffusion limited aggregation (DLA) and has also been inferred for island growth of C30 (triacontane) on silicon substrates \cite{Knufing2005FractalInterfaces}. {Above 0.06~mm/s, the fractal dimension drops reaching a minimum of $D_F \sim 1.48$ at 0.17~mm/s. The structures between 0.06 and 0.83 resemble a dragon-fly morphology. This dragonfly morphology starts stretching in the direction of the evaporation front, which in turn changes the fractal dimension for $v>$ 0.83 ~mm/s.} The fractal dimension reaches a plateau value of $D_F\sim 1.8$ at 1.66~mm/s, which correspond to more compact, stripe-like patterns. Thus, the different film deposition regimes are also discernible in the fractal dimension analysis.

 {Using AFM after sample preparation, we can analyze the film thickness of our dendritic structures. AFM images are taken in amplitude-modulation mode on a sample prepared at the lowest pulling velocity $v=$0.017~mm/s (Fig. S2) and at the highest pulling velocity $v=$3.3~mm/s (Fig. \ref{fig:dipcoating}g). By extracting color histograms from these pictures, see Fig.~\ref{fig:dipcoating}h), the step height of the molecular layer can be estimated from the distance between the histogram peaks. For the lowest  pulling velocity, we find that the step height is 3~nm, while for the highest pulling velocity we find a step height of 3.3~nm (Fig. \ref{fig:dipcoating}h). By averaging 5 AFM images at the lowest and highest pulling velocity, we obtain a step height of 3.3$\pm$0.2~nm and 3.5$\pm$ 0.1~nm, respectively. So within the error margins, the step heights are equal for the distinct velocity regimes and they correspond roughly to the 4.3\,nm of a perpendicular layer of C32 molecules. We would need a tilt angle of 44$^{o}$ of the C32 axes with regard to the surface normal to reach a full agreement between layer height and rectified alkane molecule. From the present measurements we cannot exclude such a tilt. However, there is a vast  literature, in particular x-ray scattering experiments, supporting a perpendicular layer both in stripes of alkanes as well as in the dragonfly structures that are deposited on silicon interfaces \cite{Corrales2014SpontaneousStripesb}. Therefore we rather attribute the reduced layering height to a ''false'' step height determination in the AFM measurements. This difference between our measurements and the theoretical value of 4.3 nm has been explained by a false step effect that originates when the tip goes from a parallel layer of molecules to a perpendicular layer as reported by Bai et al. \cite{Bai2008ExplanationFilms}. {Bai et al. reported that the false step height observed using amplitude modulation AFM depends on the setpoint and can be in the range of 20$\%$ of the length measured by contact mode AFM. Our measured step height ($\sim$3.3 nm) agrees with this false step observation amounting to 23$\%$ of the literature value for C32 molecules.} Therefore, we come to the conclusion that the nanoscale film structure, i.e., the molecular layering, is identical for the two coating regimes, i.e., a parallel layer followed by a perpendicular layer of molecules.}

\begin{figure}[ht]
\centering
\includegraphics[width=\textwidth]{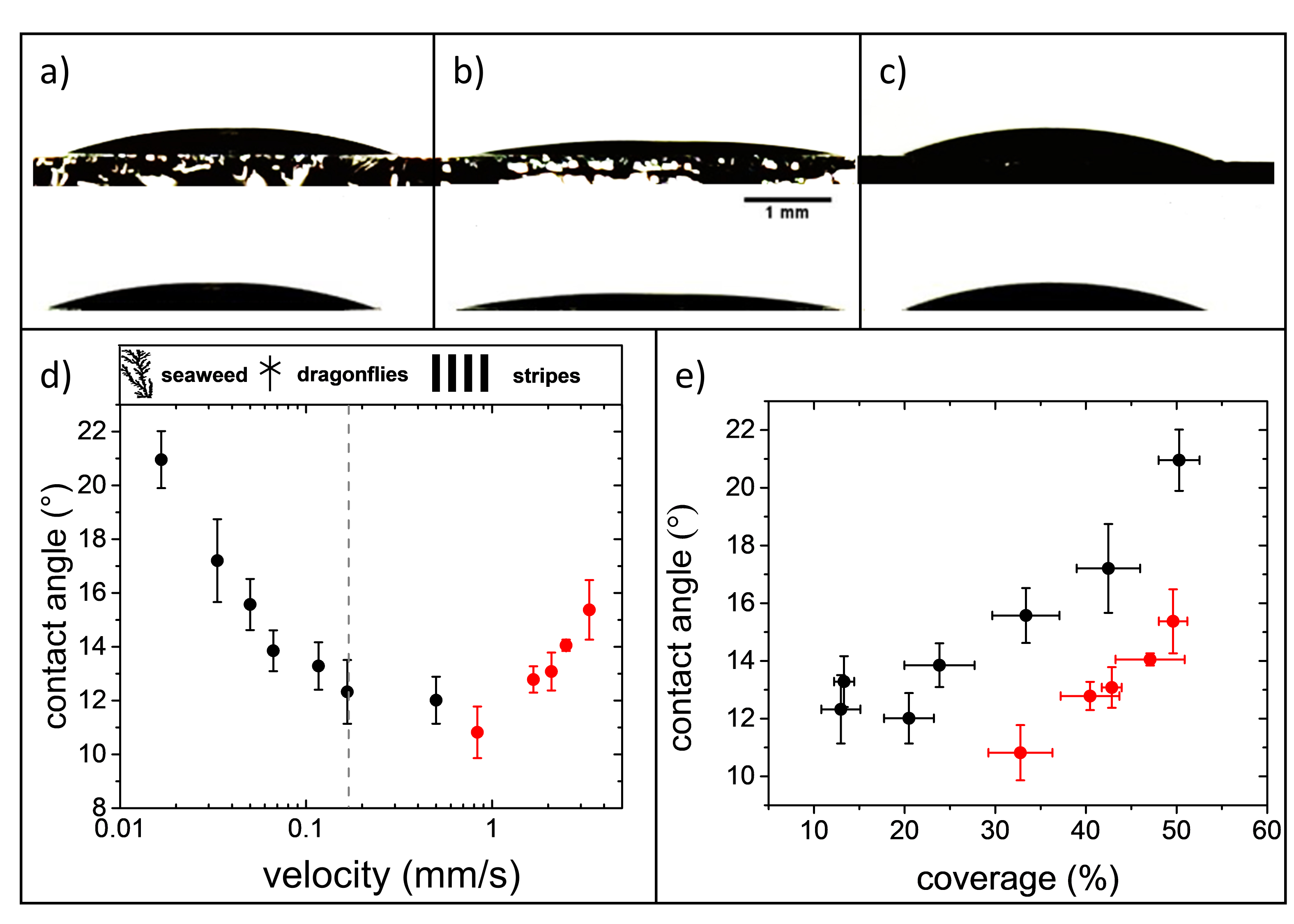}
  \caption{\textbf{Wetting properties of C32 nanopatterns.} Side-view on a water droplet sitting on a substrate prepared at a) the lowest pulling velocity (0.017~mm/s), b) at an intermediate speed (0.83~mm/s) and c) at the fastest pulling velocity 3.3~mm/s. d) Contact angle versus pulling velocity. The dashed grey line represent the critical velocity ($v_{crit}$). e) Contact angle versus surface coverage. The black and red colors indicate data points recorded from patterns formed in the evaporation and entrainment regime, respectively.}
  \label{fig:wettinganalysis}
\end{figure}

{\subsection{Exploration of wetting behaviour}}
Are these distinct 2-D nanopatterning regimes, i.e., $v<v_{crit}$ and $v>v_{crit}$, also reflected in the macroscopic wetting behaviour with water? To answer this question, contact angle measurements are performed on all samples using our home-built  sessile drop experiment. The apparent contact angle $\theta^{*}$ as a function of preparation velocity $v$ and surface coverage with the perpendicular layer are shown in Figure \ref{fig:wettinganalysis}.
The contact angle decreases with pulling velocity reaching a minimum value at a pulling velocity of 0.83 mm/s. Then it turns-over and starts increasing with pulling velocity. In Figure~\ref{fig:wettinganalysis}e we show the contact angle versus surface coverage. We note that for similar coverages, the obtained contact angles show appreciable differences, e.g at the maximum coverage ($\sim$ 50 \%) we measured  $\theta^{*}=15^\circ$ and $\theta^{*}= 22^\circ$ for low and high pulling velocities, respectively. Furthermore, given the heterogeneous surface wetting conditions we would expect a minimum in the contact angle for a minimum in C32 covered area, given the hydrophilic nature of the Si/SiO$_{\rm 2}$ substrate and the hydrophobic nature of C32. Instead, the contact angle minimum occurs at around 30$\%$ coverage. 
{
\subsection{Phenomenological wetting analysis}}
To understand why the contact angles in Figure \ref{fig:wettinganalysis}e are separated in two distinct regions{, we look at the AFM images after sample preparation and in-liquid conditions. In first place, }the distinct wetting behaviour, for identical initial hydrophobic coverage, can not be explained by differing film structures after preparation{, given that the films have the same step-height for all preparation velocities (see Fig. S2 and Fig. \ref{fig:dipcoating}f-h).  {Furthermore, we have calculated the average film height ($<h>$) and surface roughness ($R_{q}$) for all films using their measured coverage percent (see Fig. S3 and S4). We have found that the surface roughness ranges between 1.4 nm - 2.1 nm (Fig. S4) for all samples. By plotting contact angle versus roughness we find that there is no strong dependence on the initial roughness calculated from the coverage (Fig. S5). The initial film roughness varies less than 1 nm between all samples, and therefore it is unlikely to influence the macroscopic wetting properties.} Therefore, the results of Figure \ref{fig:wettinganalysis} } must be related to the distinct in-plane film morphology and the potential rearrangements of the films upon wetting with water.

\begin{figure}[h!]
\centering
\includegraphics[width=0.8\columnwidth]{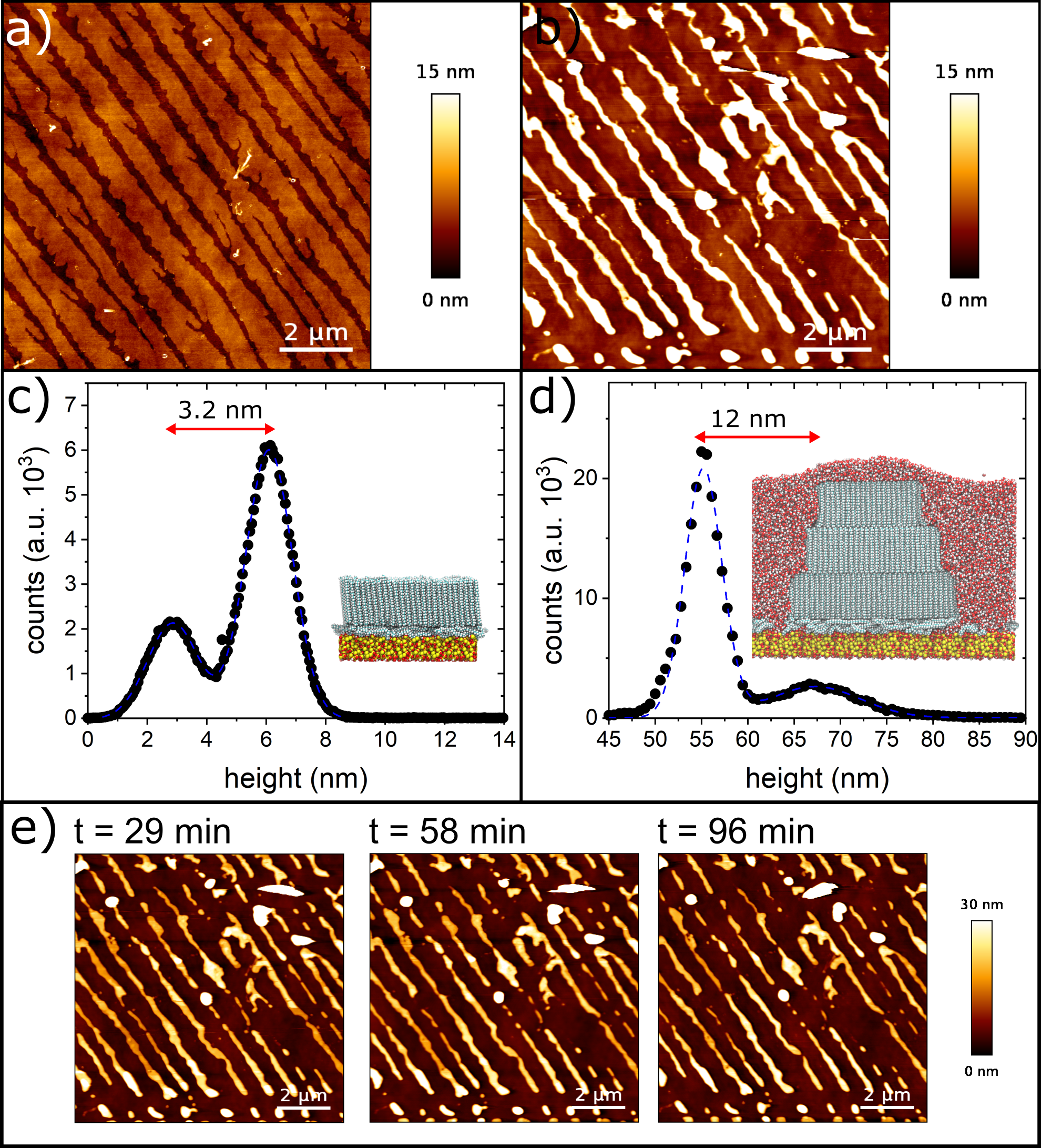}
  \caption{\textbf{In-liquid Atomic Force Microscopy (AFM) indicates water-induced reorganization of hydrocarbon nanopatterns.} a) AFM image of an as-prepared, dry sample in comparison with b) the in-liquid conditions after approximately 20 minutes along with the corresponding surface height histograms c) and d), respectively. In the as-prepared case the difference between the height histogram maxima, i.e., the difference between the dark valleys and the bright monolayer surfaces, correspond to a height typical of one perpendicular C32 monolayer, whereas in d) it corresponds to approximately 3 perpendicular C32 monolayers. e) Temporal evolution of the monlayer in liquid conditions after 29, 58 and 96 minutes. 
  \label{fig:stepheightliquid}}
\end{figure}

\begin{figure}[h!]
\centering
\includegraphics[width=0.7\textwidth]{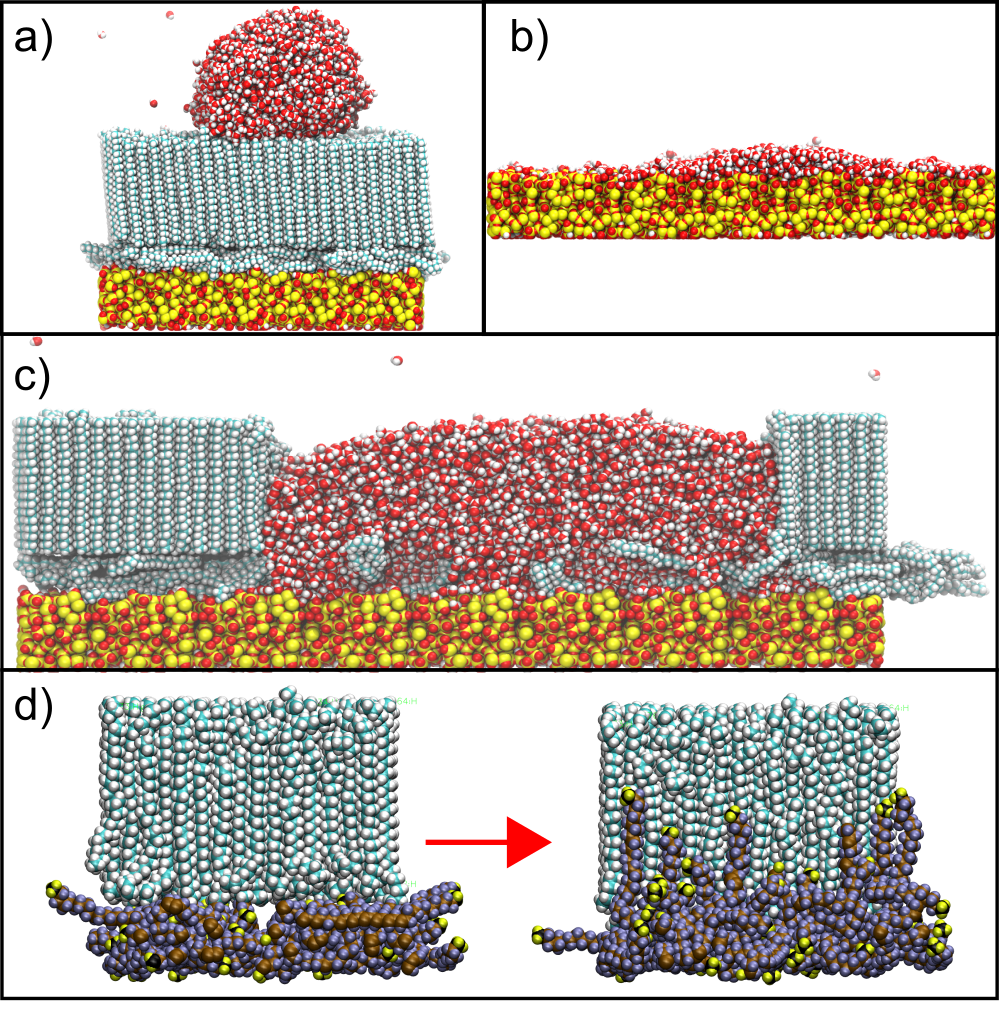}
  \caption{\textbf{Molecular Dynamics simulation snapshots of water at n-alkane/silica surfaces.} a) Drop of water on a nanopattern consisting of a perpendicular layer and a parallel bilayer of C32 molecules. b) Drop of water on a native silica layer. c) Water molecules displace the C32 parallel bilayer uncovered by a perpendicular layer. d) Parallel bilayer restructuring with C32 molecules climbing up the perpendicular layer upon contact with water.}
  \label{fgr:MDS}
\end{figure}

In Figure~\ref{fig:stepheightliquid}a we show an AFM image of a nanopattern prepared at high velocity (3.3~mm/s) in air, immediately after preparation. Fig.~\ref{fig:stepheightliquid}b shows this nanopattern within the same region while immersed in water. The nanopatterns are substantially reorganized upon the presence of water. The surface coverage with the perpendicular layer is reduced from 71\,$\upmu$m$^{2}$ to 25$\upmu$m$^{2}$, i.e. by a factor of 0.35 {over a total area of 100\,$\upmu$m$^{2}$}. Figures \ref{fig:stepheightliquid}c and d show the AFM height histograms before and after water immersion. These histograms  show an increase in height of the stripes from 3.2\,nm in air to 11.8\,nm in liquid. {The RMS roughness measured from the AFM image, for the sample prepared at 3.3~mm/s is $R_{q}=$1.8~nm in air, while underwater it increases to $R_{q}=$11.6~nm. Further AFM measurements were carried out in air and liquid conditions for samples prepared at 0.03~mm/s and 1.7~mm/s (see Figures S6 and S7). The sample prepared at 0.03~mm/s has a roughness of $R_{q}=$ 2.3~nm in air (Figure S6 left panel) and $R_{q}=$4.4 nm in liquid (Fig. S7 left panel). The sample prepared at 1.7~mm/s has a roughness of $R_{q}=$1.8~nm in air (Fig. S6 right panel) and $R_{q}=$7.4 nm in liquid (Fig. S7 right panel).} This indicates that for a quantitative analysis of the wetting behaviour, we must take into account the restructuring of the film due to the water's influence. {Furthermore, roughness measurements are indicating that structures with seaweed and dragonfly morphology change their roughness between air and liquid conditions less than  stripe-like nano-patterns.}  
To gain more mechanistic insights on what is happening at the water-nanopattern interface we perform Molecular Dynamics simulations. Figure \ref{fgr:MDS}a shows a drop of water placed over the nanopattern structure. In this configuration, the drop exhibits a high contact angle ($\theta_{\rm n-\mathrm{alkane}}\sim119^\circ$). On the other hand, a water droplet placed over a native silicon oxide completely wets the surface  (Fig. \ref{fgr:MDS}b). This means that the low, but final contact angles measured cannot be explained if the drop stands completely over the nanopattern, nor if the C32 coverage is completely removed, i.e., leaving only the native silicon.
Additional MD simulations give here important mechanistic insights. They indicate that upon water contact the molecules of the parallel bilayer, between the stripes and uncovered by a C32 bilayer, begin to migrate to the perpendicular layer (Fig. \ref{fgr:MDS}d) of the stripes on very short time scales. See the corresponding {movies in the supplementary showing a top-view of the parallel layers immersed in water (SMD1) and side-view of a water drop placed between perpendicular layers (SMD2).} It represents a MD simulation movie covering 1 ns time span of C32 rearrangement after contact with water. This migration of molecules results in an n-alkane uncovered, hydrophilic silica surface between the perpendicular layers. Moreover, {molecular climbing} explains the increased C32 terrace heights in Fig.~\ref{fig:stepheightliquid}b and d. Note that a simulation of the full water-induced alkane rearrangement would be computationally too expensive. In the view of the MD simulations, however, the measured height of 11.8~nm can be well explained by the height of 2 parallel layers + 3 perpendicular layers with respect to the bare silica substrate between the stripes. This suggests that this water-induced lateral and perpendicular mesoscopic rearrangement comes to an end as soon as 3 perpendicular layers have piled up. It involves however not only a complete removal of the uncovered C32 bilayers but also a reduction of the footprint of the perpendicular terraces until the compactified C32 coverage is inert against further rearrangements, after reaching a thickness of around 3 perpendicular monolayers.

{
We have no detailed information on the crystallographic arrangement in the perpendicular C32 layer(s) after contact with water. In our previous study we presented grazing-incidence X-ray diffraction data that indicate a herringbone-type backbone ordering in the dry perpendicular layer \cite{Corrales2014SpontaneousStripesb}. Presumably, this order prevails also in the piled-up layers and one may speculate that lattice strain energy-release limits the number of stable layer on the silica surface to roughly 3 perpendicular monolayers. It could, however, also be related to an increasingly difficult pile-up procedure and that the C32 molecules are being dissolved into the bulk water phase the further they move from the attractive silica surface.}  

In the following we will consider these important simulation insights along with the in-liquid AFM results to perform a quantitative analysis of the wetting behaviour. To model the contact angles, we first consider the classical Cassie equation for heterogeneous (hydrophilic, hydrophobic) surfaces \cite{Cassie1948ContactAngles}: 
\begin{equation}
\cos\theta^{*}=f_{\rm 1}\cos\theta_{\rm 1} + f_{\rm 2}\cos\theta_{\rm 2}  \label{eq:CassieBaxterMod}
\end{equation}
where $\theta_{\rm 1}$ is the contact angle of the C32 nanopattern and $\theta_{\rm 2}$ is the contact angle for the substrate. The terms $f_{\rm 1}$ and $f_{\rm 2}$ are the coverage of the nanopattern and native silicon oxide respectively. Considering $f_{\rm 1}+f_{\rm 2}=1$, equation \ref{eq:CassieBaxterMod} can be rewritten as:
\begin{equation}
\cos\theta^{*}=\cos\theta_{2}+(\cos\theta_{1} -\cos\theta_{2} ) f_{1}^* \alpha 
\label{eq:CassieBaxterMod alpha}
\end{equation}
where we have introduced the reduction parameter $\alpha$, which represents how much the nanopattern coverage decreases from its original value due to its interaction with the liquid interface, i.e.  $f_{\rm 1}=f_{\rm 1}^* \alpha$, where $f_{\rm 1}^*$ is the original coverage. {This means, there is more restructuring of the nanopattern as the $\alpha$ parameter becomes smaller.}

\begin{figure}[h!]
\centering
\includegraphics[width=0.8\textwidth]{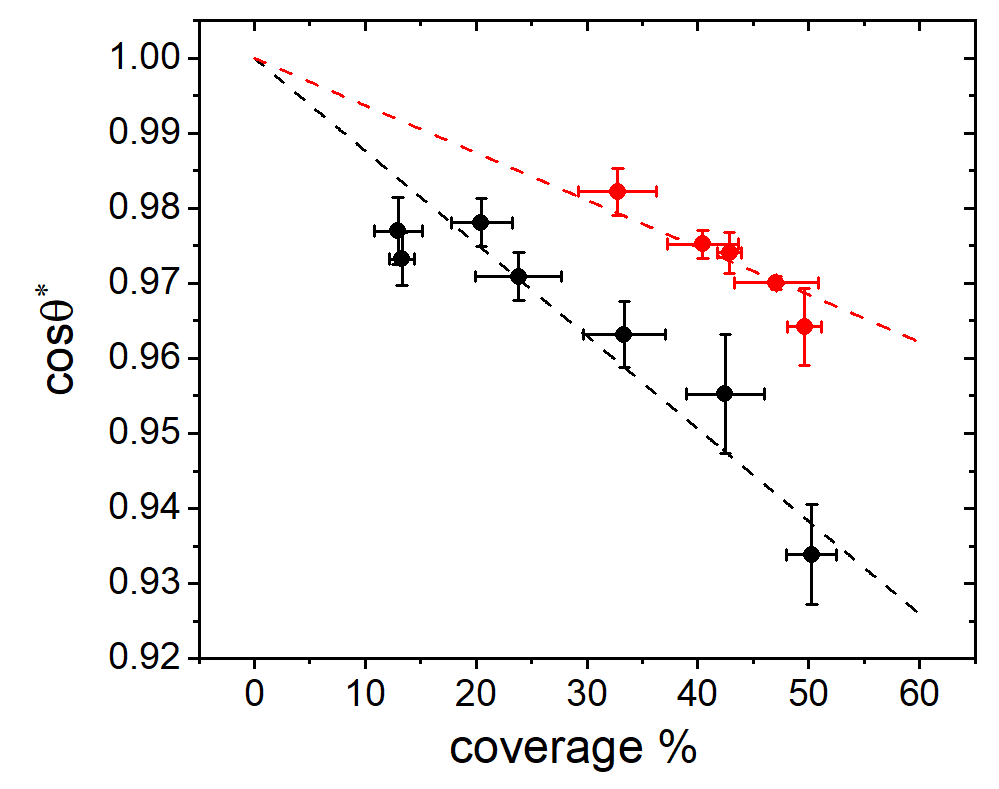}
  \caption{\textbf{Cassie wetting model analysis of distinct hydrocarbon film pattern regimes.} Cosine of apparent contact angle versus coverage. Two regimes are separated by the critical velocity $v_{\rm crit}= 0.83$~mm/s. Red dots represents the samples fabricated at $v \geq v_{\rm crit}$ and black dots for $v\leq v_{\rm crit}$. The solid lines represent fits using the modified Cassie model with the C32 coverage reduction parameter $\alpha$.}
  \label{fgr:Cassie Modified Reduction}
\end{figure}

We fit our apparent contact angles (Fig.~\ref{fig:wettinganalysis}) with the modified Cassie equation. By fixing $\theta_{n-\mathrm{alkane}}=119^{o}$ and $\theta_\mathrm{silica}=0^{o}$, as determined in our MD simulations for the homogeneous hydrophobic and hydrophilic surfaces, we obtain a reduction of $\alpha=0.08$ for samples prepared at $v<v_\mathrm{crit}$ (black points fig. \ref{fgr:Cassie Modified Reduction}) and $\alpha=0.04$ for samples prepared at $v>v_\mathrm{crit}$ (red points Fig. \ref{fgr:Cassie Modified Reduction}), respectively. Furthermore, we compare these results using a bulk C32 water contact angle of 101$^{o}$, measured on a pellet of C32. With this macroscopic contact angle we obtain a restructuring parameter of $\alpha=0.10$ for low velocities ($v<v_{\rm crit}$) and $\alpha=0.05$ for samples prepared at high velocities($v>v_{\rm crit}$). The effect of the two possible contact angles for C32 lead to similar results when comparing the two types of morphology.
From the model presented in Fig.~\ref{fgr:Cassie Modified Reduction} we conclude that nanopatterns prepared at low dip-coating velocities restructure less than samples prepared at higher velocities. This could be due to the intrinsic morphology of the two regimes. The patterns prepared at low velocities are less compact than the patterns prepared at higher velocities. {Therefore, the water molecules may not be able to enter the highly tortuous, dendritic pathways and 2-D niches towards the inner parts of the C32 carpets, given the hydrophobic boundaries. By contrast, for the compact stripes with better defined 2-D boundaries, water may be able to reach the C32 molecules more easily and thus can presumably coherently displace entire rows of C32 molecules, also since the transport pathways are less meandering than in the case of the dendritic 2-D C32 crystals. However, this hypothesis has to be explored in more detailed manner in future studies. In particular, extensive MD simulations could be helpful to achieve a mechanistic understanding of this distinct behaviour.}
{\subsection{Surface thermodynamics}}
{
The microscopic interfacial structure and its time evolution is overall quite complex. Nevertheless surface thermodynamics should provide mechanistic insights on the causes for the restructuring process and the resulting film structure. According to wetting thermodynamics the interface strives torward a situation with minimal surface energy. Therefore three cases in terms of layering arrangement have to be compared with respect to their energetics and the sum of the corresponding interfacial energies: (i) the bare silica-vapour surface, (ii) the hydrated silica/C32/water/vapour and (iii) the non-hydrated silica/C32/water/vapour layering. Values for the corresponding interfacial energies are reported in the literature for the bare amorphous silica-vapour,$\gamma_{\rm Si0_2/v} =260\pm$13~mN/m \cite{Brunauer1956}, the hydrated amorphous silica-vapour, $\gamma_{\rm h-Si0_{2}/v}=130 \pm$ 8~mN/m \cite{Brunauer1956}, the C32-water, $\gamma_{\rm C32/w}=50\pm$5~mN/m \cite{Kramer2019}, the water-vapour $\gamma_{\rm w/v}$=72$\pm$5~mN/m \cite{deGennes2004CapillarityPhenomena} and the C32-vapour interfaces $\gamma_{\rm C32/v}$=30$\pm$3 mN/m \cite{Ocko1997}, respectively. For the theoretical value of the contact angle of water on bulk C32 this results, according to the Young-Laplace equation in $\theta_{n-\mathrm{alkane}}=104^{o}$, a value in good agreement with the measured 101$^{o}$ for a bulk pellet of C32. Moreover, the difference between the hydrated and unhydrated silica surface energies, $\gamma_{\rm Si0_2/v}-\gamma_{\rm h-Si0_2/v} \sim$ 130~mN/m indicate a substantial reduction in excess free energy upon silica hydration. Unfortunately, we are not aware of any value reported for the C32/amorphous silica interfacial energy. However based on measured contact angles for medium-length n-alkanes on amorphous silica of about 30$^{o}$ \cite{Ingram1974} we estimate it to be $\gamma_{\rm C32/Si0_2}\sim$ 230~mN/m, significantly larger than $\gamma_{\rm h-SiO_2}$. Hence, in agreement with our MD simulations the surface thermodynamics indicates that water is preferred at the silica surface in comparison to C32. Moreover, the film structure consisting of hydrated silica-C32-air results in an overall surface excess energy of $\gamma_{\rm h-Si0_2/v}+\gamma_{\rm C32/w}+ \gamma_{\rm C32/v}$ of $\sim$ 210 mN/m, well below the initial bare silica surface of 260 mN/m and lower than the estimated excess energy for a layering silica/C32/water of $\sim$ 350 mN/m. Thus, the observed film evolution agrees qualitatively and semi-quantitatively with simple phenomenological surface thermodynamics provided one neglects the heterogeneous wetting geometry and the details of the microscopic inhomogenoeus film structures.
}
{
These energetic considerations also suggest that the observed physisorbed C32 layers may be rather metastable, observable only on the exploration times of hours after contact with water, as typical for our experiments. Eventually, a complete removal of physisorbed C32 layers from the silica substrates would be the energetically and thus thermodynamically most favoured equilibrium state. This highlights the peculiar self-hydrophilisation tendency of water at hydroxylated silica surfaces with respect to physisorbed hydrocarbons that eventually should lead to perfect wetting of the initially rather hydrophobic surface coatings.}
\\
\hl{Finally, we also added SEM images of a sample prepared at 0.033 mm/s after complete evaporation of the droplet (Figure S8). From these images we can identify three regions: an exterior,  intermediate and central region. The exterior region is unaffected by the liquid droplet and exhibits sea-weed like morphology. The central region exhibits a nano-pattern that resembles our AFM results under liquid conditions. This means the nano-patterns has a seaweed-like morphology, but is thinner than the nano-patterns in the exterior region. The intermediate region exhibits a nano-pattern that is completely restructured. Remarkably, there is also a circular region that is completely devoid of adsorbed hydrocarbons. Presumably, the intermediate region result from a complex interplay of alkane dissolution in water, evaporation-induced complex flow patterns at the contact line and substantial shear flows in the surface proximity during the drying. This intermediate region goes beyond the scope of the present study and deserves further experimental and theoretical studies in the future. It is, however, reminiscent of the complex material re-deposition processes known from droplet drying of liquid suspensions, most prominently from the coffee ring effect \cite{Deegan1997, Ristenpart2007}.}
\\

\section{Conclusions}
We have studied the wetting properties of self-organised n-alkane nanopatterns prepared by dip-coating. We show that both the coverage and morphology of the nanopatterns can be controlled by the sample preparation velocity. Furthermore, we find a coating transition at 0.83\,mm/s between two types of organic growth structure, which is reflected in both the surface coverage as well as the fractal dimension of the resulting nanopatterns. The transition between the two coating regimes affects the wetting properties of water, i.e., the measured contact angles also group into two regimes. 

By using {complementary} Atomic Force Microscopy and Molecular Dynamics simulations we see that the nanopatterns are not completely stable. The water layer is capable of removing  the parallel bilayer that is uncovered by the perpendicular layer, as well as the coverage consisting of perpendicular layers, leading to the formation of terraces with up to 3 perpendicular monolayers. This water-induced restructuring results in a substantial reduction of the hydrophobic surface coverage. Based on these mesoscopic insights on the molecular rearrangement of the hydrophobic alkane molecules, we have quantified the resulting hydrophilization process using a modified Cassie model. It incorporates a reduction parameter that quantifies the observed restructuring of the nanopatterns. We use the contact angles seen in simulations as fixed parameters in the modified Cassie model and obtained a reduction parameter for the two contact angle regimes. We deduce from this model that the reduction parameter is larger for samples prepared below the critical velocity than samples prepared at higher velocities. We can correlate this with the morphology of the nanopatterns. Patterns with dendritic shape, as prepared in the evaporation regime, restructure {less} than samples prepared at higher velocity in the entrainment regime, with a more compact stripe appearance. Thus, we propose that the lateral roughness of the nanopattern can influence the effective hydrocarbon displacement capacity of the water molecules.

Given the omnipresence of physisorption of hydrocarbons from the environment, in particular from airborne sources \cite{Li2013,Li2016, Terzyk2019}, our study evidences that water is nevertheless able to reduce the resulting surface hydrophobicity, without necessarily dissolving the hydrocarbons into the bulk liquid. Rather the alkanes remain in a physisorbed state, rearranging into more compact structures encompassing multilayer terraces. {A surface thermodynamic analysis also suggests that these rearranged physisorbed layers are rather metastable, i.e., stable on our experimental exploration times of hours and days after contact with water. The energetically most favourable state is a complete removal of the physisorbed alkane layers from the hydroxylated silicon surfaces.}

\hl{There are several publications presenting experiments, theoretical considerations and computer simulations that indicate a similar monolayer structure for physisorbed medium- to long-length n-alkanes on silicon surfaces with a native oxide layer (15\r{A}) \cite{Tanaka2008,merkl1997,Mo2004StructureSimulations,Scholllmeyer2003} and even with a thick silicon oxide layer (300 nm) \cite{Koehler2006}. These studies cover chain lengths from 16 to 50 carbon atoms (C16-C50) deposited over a silica interface, all showing a similar molecular layering as our system. Therefore, we are encouraged to believe that for these systems a similar interplay of intermolecular and surface interactions occurs and thus analogous self-hydrophilisation effects can be expected upon contact with water.}

This observation is of particular relevance for interface-dominated structures and thus for processes in nanoporous media or at nanostructured surfaces, where the hydrocarbon displacement mechanism explored here at planar surfaces acts like a self-cleaning mechanism and could substantially alter self-diffusion dynamics \cite{Jani2021} and hydraulic permeabilities \cite{Gruener2009, Kusmin2010a, Huber2015, Gruener2016, Gruener2016a} upon water transport {and phase transitions \cite{Gang2020}}. Specifically for capillarity-driven, spontaneous imbibition, it also means that by water wetting a quite effective displacement of hydrocarbons should be possible, in agreement with recently reported experiments on water-condensation induced oil-displacement in mesoporous silicas \cite{Gimenez2019}.

\section*{Conflicts of interest}
There are no conflicts to declare.

\section*{Acknowledgements}
UGV acknowledges support of Fondecyt 1180939. TPC acknowledges the support of Fondecyt Iniciación 11160664, PCI MPG190023. PH acknowledges support by the Deutsche Forschungsgemeinschaft (DFG) within the priority program SPP 2171, ''Dynamic wetting of flexible, adaptive and switchable surfaces'', Projektnummer 22879465. RM acknowlegdes support by the Deutsche Forschungsgemeinschaft (DFG, German Research Foundation) -- Projektnummer 192346071 -- SFB 986 and -- Projektnummer 390794421 -- GRK 2462. PH and RM profited also from Grant-No. HIDSS-0002 DASHH (Data Science in Hamburg - HELMHOLTZ Graduate School for the Structure of Matter). We also acknowledge the scientific exchange and support of the Center for 
Molecular Water Science (CMWS).


\end{document}